\documentclass[twocolumn, showpacs, preprintnumbers, nofootinbib, aps, prd, superscriptaddress, 10pt, showkeys, notitlepage]{revtex4-1}

\usepackage[utf8]{inputenc}
\usepackage{graphicx,amssymb,amsmath,amsthm,amsfonts,epsfig,times,natbib}
\usepackage[linktocpage]{hyperref}
\usepackage[usenames,dvipsnames]{color}
\usepackage{epstopdf}
\usepackage{aas_macros,amsmath,amssymb}
\usepackage{tensor}
\usepackage{mathtools}
\usepackage{amsbsy}
\usepackage{bm}

\definecolor{oxfordblue}{rgb}{0.0, 0.13, 0.28}
\definecolor{burgundy}{rgb}{0.5, 0.0, 0.13}
\definecolor{crimsonglory}{rgb}{0.75, 0.0, 0.2}
\definecolor{darkolivegreen}{rgb}{0.33, 0.42, 0.18}
\definecolor{darkblue}{rgb}{0,0,0.5}
\definecolor{richcarmine}{rgb}{0.84, 0.0, 0.25}
\definecolor{darkblue}{rgb}{0,0,0.5}
\definecolor{venetianred}{rgb}{0.78, 0.03, 0.08}
\definecolor{skobeloff}{rgb}{0.0, 0.48, 0.45}
\hypersetup{colorlinks=true, citecolor=darkblue, linkcolor=darkblue,
urlcolor = darkblue}

\newcommand{\be}{\begin{equation}}
\newcommand{\ee}{\end{equation}}
\newcommand{\bear}{\begin{eqnarray}}
\newcommand{\eear}{\end{eqnarray}}

\newcommand{\minf}{M_{\infty}}

\newcommand{\cP}{{\cal P}}

\newcommand{\cM}{{\cal M}}

\begin{document}
\title{Astrophysical applications of the post-Tolman-Oppenheimer-Volkoff formalism}

\begin{abstract}
  The bulk properties of spherically symmetric stars in general
  relativity can be obtained by integrating the
  Tolman-Oppenheimer-Volkoff (TOV) equations.  In previous
  work~\cite{Glampedakis:2015sua} we developed a “post-TOV” formalism
  -- inspired by parametrized post-Newtonian theory -- which allows us
  to classify in a parametrized, phenomenological form all possible
  perturbative deviations from the structure of compact stars in
  general relativity that may be induced by modified gravity at second
  post-Newtonian order. In this paper we extend the formalism to deal
  with the stellar exterior, and we compute several potential
  astrophysical observables within the post-TOV formalism: the surface
  redshift $z_s$, the apparent radius $R_{\rm app}$, the Eddington
  luminosity at infinity $L_{\rm E}^\infty$ and the orbital
  frequencies. We show that, at leading order, all of these quantities
  depend on just two post-TOV parameters $\mu_1$ and $\chi$, and we
  discuss the possibility to measure (or set upper bounds on) these
  parameters.
\end{abstract}

\author{Kostas Glampedakis}
\email{kostas@um.es}
\affiliation{Departamento de F\'isica, Universidad de Murcia,
Murcia, E-30100, Spain}
\affiliation{Theoretical Astrophysics, University of T\"ubingen,
T\"ubingen, D-72076, Germany}

\author{George Pappas}
\email{gpappas@olemiss.edu}
\affiliation{Department of Physics and Astronomy,
The University of Mississippi, University, MS 38677, USA}

\author{Hector O. Silva}
\email{hosilva@phy.olemiss.edu}
\affiliation{Department of Physics and Astronomy,
The University of Mississippi, University, MS 38677, USA}

\author{Emanuele Berti}
\email{eberti@olemiss.edu}
\affiliation{Department of Physics and Astronomy,
The University of Mississippi, University, MS 38677, USA}
\affiliation{Departamento de F\'isica, CENTRA, Instituto Superior
T\'ecnico, Universidade de Lisboa, Avenida Rovisco Pais 1,
1049 Lisboa, Portugal}

\date{{\today}}


\maketitle


\section{Introduction}
\label{sec:intro}

Compact stars are ideal astrophysical environments to probe the
coupling between matter and gravity in a high-density, strong gravity
regime not accessible in the laboratory. Cosmological observations and
high-energy physics considerations have spurred extensive research on
the properties of neutron stars, whether isolated or in binary
systems, in modified theories of gravity (see
e.g.~\cite{Berti:2015itd,Yunes:2013dva} for reviews). Different extensions of
general relativity (GR) affect the bulk properties of the star (such
as the mass $M$ and radius $R$) in similar ways for given assumptions
on the equation of state (EOS) of high-density matter. Therefore it is
interesting to understand whether these deviations from the
predictions of GR can be understood within a simple parametrized
formalism.  The development of such a generic framework to
understand how neutron star properties are affected in modified
gravity is even more pressing now that gravitational-wave observations
are finally a reality~\cite{Abbott:2016blz}, since the observation of
neutron star mergers could allow us to probe the dynamical behavior of
these objects in extreme environments.

In previous work we developed a post-Tolman-Oppenheimer-Volkoff
(henceforth, post-TOV) formalism valid for spherical
stars~\cite{Glampedakis:2015sua}.  The basic idea is quite simple. The
structure of nonrotating, relativistic stars can be determined by
integrating two ordinary differential equations: one of these
equations gives the ``mass function'' and the other equation -- a
generalization of the hydrostatic equilibrium condition in Newtonian
gravity -- determines the pressure profile and the stellar radius,
defined as the point where the pressure vanishes.  The post-TOV
formalism, reviewed in Section~\ref{sec:summary} below, adds a
relatively small number of parametrized corrections with parameters
$\mu_i$ ($i=1, \dots\,, 5$) and $\pi_i$ ($i=1, \dots\,, 4$) to the
mass and pressure equations.
These corrections have two properties: (i) they are of second
post-Newtonian (PN) order, because first-order deviations are already
tightly constrained by observations; and (ii) they are general enough
to capture in a phenomenological way all possible deviations from the
mass-radius relation in GR.
Other parametrizations were explored in~\cite{Schwab:2008ce,Velten:2016bdk}
by modifying ad hoc the TOV equations.

In this paper we turn to the investigation of astrophysical
applications of the formalism. Part of our analysis is inspired by
previous work by Psaltis~\cite{Psaltis:2007rv}, who showed that, under
the assumption of spherical symmetry, many properties of neutron stars
in metric theories of gravity can be calculated using only
conservation laws, Killing symmetries, and the Einstein equivalence
principle, without requiring the validity of the GR field
equations. Psaltis computed the gravitational redshift of a surface
atomic line $z_s$, the Eddington luminosity at infinity
$L_{\rm E}^\infty$ (thought to be equal to the touchdown luminosity of
a radius-expansion burst), and the apparent surface area of a neutron
star (which is potentially measurable during the cooling tails of
bursts).

We first extend our previous work to study the exterior of neutron
stars. Then we compute the surface redshift $z_s$, the apparent radius
$R_{\rm app}$ and the Eddington luminosity at infinity $L_{\rm E}^\infty$.
In addition, we study geodesic motion in the neutron star spacetime within the post-TOV formalism. We
focus on the orbital and epicyclic frequencies, that according to some models --
such as the relativistic precession
model~\cite{StellaApJ98,StellaPRL99} and the epicyclic resonance
model~\cite{Abramowicz:2001bi} -- may be related with the
quasi-periodic oscillations (QPOs) observed in the X-ray spectra of
accreting neutron stars.

Our main result is that, at leading order, all of these quantities
depend on just \textit{two} post-TOV parameters: $\mu_1$ and the combination
\be
\chi \equiv \pi_2 - \mu_2 - 2\pi \mu_1\,.
\label{eq:chi_def}
\ee
We also express the leading multipoles in a multipolar expansion of the
neutron star spacetime in terms of $\mu_1$ and $\chi$, and we discuss
the possibility to measure (or set upper bounds on) these parameters
with astrophysical observations.

The plan of the paper is as follows. In Section~\ref{sec:summary} we
present a short review of the post-TOV formalism developed
in~\cite{Glampedakis:2015sua}. In Section~\ref{sec:metric} we extend
the formalism to deal with the stellar exterior, computing a
``post-Schwarzschild'' exterior metric.
In Section~\ref{sec:redshift} we compute the surface redshift $z_s$
and relate it to the stellar compactness $M/R$.
In Section~\ref{sec:burst}, following~\cite{Psaltis:2007rv}, we study
the properties of bursting neutron stars in the post-TOV framework.
In Section~\ref{sec:qpos} we calculate the orbital frequencies.  In
Section~\ref{sec:multipoles} we look at the leading-order multipoles
of post-TOV stars. Then we present some conclusions and possible
directions for future work.


\section{Overview of the post-TOV formalism}
\label{sec:summary}

Let us begin with a review of the post-TOV formalism introduced in
\cite{Glampedakis:2015sua}. The core of this formalism consists of the
following set of ``post-TOV" structure equations for static spherically
symmetric stars (we use geometrical units $G=c=1$):
\begin{subequations}
\label{eq:PTOV_2PN_intro}
\begin{align}
\frac{dp}{dr} &= \left (\frac{dp}{dr} \right )_{\rm  GR}  -\frac{\rho
                m}{r^2} \left (\, {\cal P}_1 + {\cal P}_2\, \right ),
\label{eq:PTOV_2PN2_dp}
\\
\nonumber \\
\frac{dm}{dr} & = \left ( \frac{dm}{dr} \right )_{\rm GR} + 4\pi r^2
                \rho \left ( {\cal M}_1 + {\cal M}_2\right ),
\label{eq:PTOV_2PN2_dm}
\end{align}
\end{subequations}
where
\begin{subequations}
\label{eq:PandM}
\begin{align}
{\cal P}_1 &\equiv \delta_1 \frac{m}{r} + 4\pi \delta_2  \frac{r^3 p}{m},
\quad
{\cal M}_1 \equiv   \delta_3 \frac{m}{r} + \delta_4 \Pi,
\\
\nonumber\\
{\cal P}_2 &\equiv \pi_1 \frac{m^3}{r^5\rho} + \pi_2 \frac{m^2}{r^2}
+ \pi_3 r^2 p + \pi_4 \frac{\Pi  p}{\rho},
\label{eq:P_2}
\\
\nonumber\\
{\cal M}_2 &\equiv \mu_1 \frac{m^3}{r^5\rho} + \mu_2 \frac{m^2}{r^2}
 + \mu_3 r^2 p
+ \mu_4 \frac{\Pi  p}{\rho} + \mu_5 \Pi^3 \frac{r}{m} . \nonumber \\
\label{eq:M_2}
\end{align}
\end{subequations}
Here $r$ is the circumferential radius, $m$ is the mass function, $p$
is the fluid pressure, $\rho$ is the baryonic rest mass density,
$\epsilon$ is the total energy density and $\Pi \equiv (\epsilon-\rho)/\rho$ is the
internal energy per unit baryonic mass. A ``GR'' subscript denotes the standard TOV equations in GR, i.e.
\begin{subequations}
\label{eq:TOV}
\begin{align}
& \left (\frac{dp}{dr} \right )_{\rm GR} = -\frac{(\epsilon + p)}{r^2}  \frac{ (m_{\rm T} + 4\pi r^3 p  )}{  ( 1-2m_{\rm T} /r )},
\label{dpdr_tov}
\\
\nonumber \\
& \left ( \frac{dm}{dr} \right)_{\rm GR} = \frac{d m_{\rm T}}{dr} =  4\pi r^2 \epsilon,
\label{dmdr_tov}
\end{align}
\end{subequations}
where $m_{\rm T}$ is the GR mass function.

The dimensionless combinations $\cP_1,\cM_1$ and $\cP_2, \cM_2$
represent a parametrized departure from the GR stellar structure and
are linear combinations of 1PN- and 2PN-order terms, respectively.
These terms feature the phenomenological post-TOV parameters
$\delta_i$ ($i=1,\,\dots \,,4$), $\pi_i$ ($i=1,\,\dots \,,4$) and
$\mu_i$ ($i=1, \dots\,, 5$).
In particular, the coefficients $\delta_i$ attached to the 1PN terms
are simple algebraic combinations of the traditional PPN parameters
$\delta_1 \equiv  3 (1+ \gamma) -6\beta + \zeta_2$,
$\delta_2 \equiv  \gamma-1+\zeta_4$,
$\delta_3 \equiv -\frac{1}{2} \left ( 11 + \gamma -12\beta + \zeta_2 -2\zeta_4 \right )$,
$\delta_4\equiv \zeta_3.$
As such, they are constrained to be very close to zero by existing Solar System
and binary pulsar observations\footnote{Using the latest constraints on the PPN
parameters \cite{Will:2014xja} we obtain the following upper limits:
$| \delta_1| \lesssim 6\times 10^{-4}, |\delta_2| \lesssim 7\times
10^{-3}, | \delta_3 | \lesssim 7\times 10^{-3}, |\delta_4| \lesssim
10^{-8}$.}:
$|\delta_i| \ll 1$. This result translates to negligibly small 1PN terms in
Eq.~(\ref{eq:PTOV_2PN_intro}): ${\cal P}_{1}\ll1$, $\cM_1 \ll1$.
On the other hand, $\pi_i$ and $\mu_i$ are presently unconstrained, and
consequently $\cP_2, \cM_2$ should be viewed as describing the dominant
(significant) departure from GR.
The GR limit of the formalism corresponds to setting all of these parameters to zero, i.e. $\delta_i, \pi_i, \mu_i \to 0$.

Alternatively, the stellar structure
equations~(\ref{eq:PTOV_2PN_intro}) can be formally derived -- if we
neglect the small terms $\cP_1, \cM_1$ -- from a covariantly conserved
perfect fluid stress energy tensor~\cite{Glampedakis:2015sua}:
\be
\nabla_\nu T^{\mu\nu} = 0, \qquad T^{\mu\nu} = (\epsilon_{\rm eff} + p) u^\mu u^\nu + p g^{\mu\nu},
\ee
where the effective, gravity-modified energy density is
\be
\epsilon_{\rm eff} = \epsilon + \rho {\cal M}_2,
\label{eq:effectiveEOS}
\ee
and the covariant derivative is compatible with the effective post-TOV metric
\be
g_{\mu\nu} = \mbox{diag} [\, -e^{\nu(r)}, (1-2m(r)/r)^{-1}, r^2, r^2 \sin^2\theta\,],
\ee
with
\be
\frac{d\nu}{dr} = \frac{2}{r^2\\} \left [\, (1-{\cal M}_2) \frac{m+4\pi r^3 p}{1-2m/r} + m\cP_2 \, \right ].
\label{eq:nu_int}
\ee
This post-TOV metric is valid in the {\it interior} of the star. In the following section we discuss
how an \textit{exterior} post-TOV metric can be constructed within our framework.


\section{The exterior ``post-Schwarzschild'' metric}
\label{sec:metric}

For the applications of the post-TOV formalism considered in this work, we must
specify how the $g_{tt}$ and $g_{rr}$ metric elements are calculated in the interior
and exterior regions of the fluid distribution. In this section we will construct
an exterior spacetime in a ``post-Schwarzschild'' form.

From the effective post-TOV metric, we have that {\it inside} the
fluid body $g_{tt} = -\exp[\nu(r)]$, where $\nu(r)$ is determined in
terms of the fluid variables and post-TOV parameters from
Eq.~(\ref{eq:nu_int}). We will assume that {\it outside} the fluid
distribution the {\it same} effective metric expression
holds\footnote{This assumption is based on simplicity. While we keep
  an agnostic view on the validity of Birkhoff's theorem within our
  formalism (and in modified gravity theories in general), the
  interior post-TOV metric is arguably the best guide towards the
  construction of the exterior metric.  }.
Then we get the equations
\bear
&& \frac{d\nu}{dr} = \left (\frac{d\nu}{dr} \right )_{\rm GR}  +  \frac{2}{r^2} \left [\, - \mu_2 \frac{m^2}{r^2} \frac{m}{1-2m/r}
 + \pi_2 \frac{m^3}{r^2} \, \right ],
 \label{dnu_ext0}
 \nonumber \\
\\
\nonumber \\
&& \frac{dm}{dr} =  4\pi  \mu_1 \frac{m^3}{r^3},
\label{dm_ext1}
\eear
where
\be
\left (\frac{d\nu}{dr} \right )_{\rm GR} \equiv  \frac{2}{r^2}
\frac{m}{1-2m/r}.
\label{eq:nu_ext_gr}
\ee
These equations originate from the general expressions
(\ref{eq:nu_int}) and (\ref{eq:PTOV_2PN2_dm}) after setting all fluid
parameters to zero, i.e $p=\epsilon=\rho=\Pi=0$, and keeping the
surviving terms in $\cM_2$ and $\cP_2$.
It is not difficult to see that, in the nomenclature of \cite{Glampedakis:2015sua},
the only 2PN post-TOV terms that can appear in the exterior
equations are those of ``family F1''  and ``family F2''.
 The F1 term (coefficient $\pi_1$) should not appear in the $\cP_2$ correction of
the \textit{interior} $d\nu/dr$ equation because it is divergent at the surface.
This implies that the F1 term should not appear in the $dp/dr$ equation either.

As it stands, Eq.~(\ref{dnu_ext0}) contains higher than 2PN order terms.
It should therefore be PN-expanded with respect to the post-TOV terms:
\be
\frac{d\nu}{dr} = \left (\frac{d\nu}{dr} \right )_{\rm GR}  +  2( \pi_2 - \mu_2) \frac{m^3}{r^4}.
 \label{dnu_ext1}
\ee
Thus  (\ref{dnu_ext1}) and (\ref{dm_ext1}) are our ``final'' post-TOV equations for the stellar exterior.

The mass equation is decoupled and can be directly integrated. The result is
\be
m(r) = \frac{r}{\sqrt{4\pi \mu_1 + K r^2}},
\ee
where $K$ is an integration constant. The fact that $dm/dr \neq 0$ outside the star
 implies the presence of an ``atmosphere'' due to the non-GR degree of freedom.
This is reminiscent of the exterior structure of neutron stars in
scalar-tensor theories \cite{Damour:1993hw}. The constant $K$ is fixed by setting
$m(r\to\infty) $ equal to the system's ADM mass $\minf$. Then,\
\be
m(r) = M_\infty \left ( 1+  4\pi \mu_1 \frac{M^2_\infty}{r^2} \right )^{-1/2}.
\label{m_ext}
\ee
Thus the ADM mass is related to the Schwarzschild mass $M \equiv m(R)$ by
\be
M = M_\infty \left ( 1+  4\pi \mu_1 \frac{M^2_\infty}{R^2} \right )^{-1/2}.
\label{eq:m_inf0}
\ee
As expected, in the GR limit the two masses coincide
\be
m(r> R) = M_\infty = M.
\ee
Assuming a post-TOV correction ${\cal F}  \equiv 4\pi |\mu_1 | (M_\infty/R)^2 \ll 1$ we can
re-expand our result~(\ref{eq:m_inf0}),
\be
M = M_\infty \left ( 1-  2\pi \mu_1 \frac{M^2_\infty}{R^2} \right ).
\label{eq:m_inf1}
\ee
The inverse relation $M_\infty = M_\infty (M)$ reads\footnote{At first
  glance, obtaining $M_\infty (M)$ entails solving a cubic equation.
  However, the procedure is greatly simplified if we recall that the
  post-TOV formalism must reduce to GR for $\{\mu_i, \pi_i\} \to
  0$.
  Having that in mind we can treat $\mu_1$ as a small parameter and
  solve (\ref{eq:m_inf1}) perturbatively.  The only regular solution
  in the $\mu_1 \to 0$ limit is Eq.~(\ref{eq:m_inf2}).}
\be
M_\infty = M \left ( 1+  2\pi \mu_1 \frac{M^2}{R^2} \right ).
\label{eq:m_inf2}
\ee
The three mass relations~(\ref{eq:m_inf0}), (\ref{eq:m_inf1}) and
(\ref{eq:m_inf2}) are equivalent in the ${\cal F} \ll 1$
limit. Equations~(\ref{eq:m_inf0}) and (\ref{m_ext}) are ``exact"
post-TOV results and do not require ${\cal F}$ or $\mu_1$ to be much
smaller than unity, although Eq.~(\ref{eq:m_inf0}) does place a lower
limit on $\mu_1$ because the argument of the square root must be
nonnegative. Unfortunately, the use of (\ref{m_ext}) in the
calculation of the metric components leads to very cumbersome
expressions.

\begin{figure}[htb]
\includegraphics[width=0.47\textwidth]{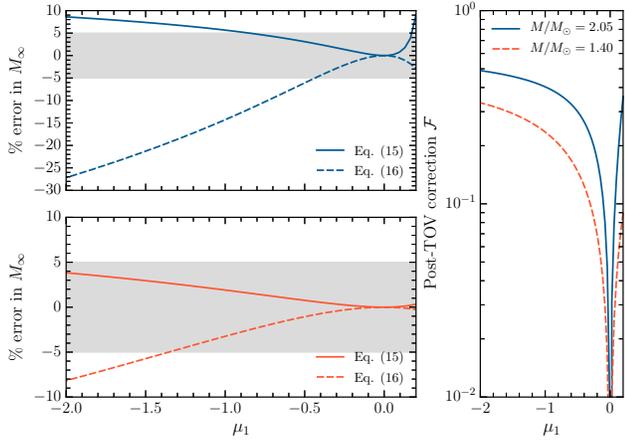}
\caption{{\it Errors in $M_{\infty}$}.
We show the percent error [$\%\, {\rm error} \equiv 100 \times (x_{\rm value} - x_{\rm ref})/x_{\rm ref}$]
in calculating $M_{\infty}$ using Eqs.~(\ref{eq:m_inf1}) and (\ref{eq:m_inf2})
with respect to (\ref{eq:m_inf0}) for various values of $\mu_1$ using EOS SLy4.
The range of $\mu_1$ is chosen such that using any of Eqs. (\ref{eq:m_inf0}), (\ref{eq:m_inf1})
or (\ref{eq:m_inf2}) one can obtain a real root corresponding to $M_{\infty}$.
The post-TOV models are constructed using a fixed central value of the energy density,
which results in either a canonical ($1.4\, M_{\odot}$) or a maximum-mass
 ($2.05\, M_{\odot}$) neutron star in GR.
{\it Top panel:} errors for a maximum-mass GR star.
%
{\it Bottom panel:} errors for a canonical-mass GR star.
%
{\it Right panel:} the absolute value of the post-TOV correction
${\cal F} =4 \pi \mu_1 (M_{\infty}/R)^2$ as a function of $\mu_1$. The
condition ${\cal F} \ll 1$ bounds
the range of acceptable values of $\mu_1$ for which the expansions
leading to
Eqs.~(\ref{eq:m_inf1}) and (\ref{eq:m_inf2}) are valid.
Errors are below 5 $\%$ when $\mu_1 \in [-1.0, 0.1]$.  }
\label{fig:errors}
\end{figure}

To make progress (while also keeping up with the post-TOV spirit), we hereafter use the ${\cal F} \ll 1$
approximations (\ref{eq:m_inf1})-(\ref{eq:m_inf2}). This step, however, introduces a certain degree of error.
This is quantified in Fig.~\ref{fig:errors} (left panel), where we show the relative percent error in calculating $M_{\infty}$
using the post-TOV expanded Eqs.~(\ref{eq:m_inf1})
and~(\ref{eq:m_inf2}) rather than Eq.~\eqref{eq:m_inf0}.
Using EOS Sly4~\cite{Douchin:2001sv}, we considered values of $\mu_1$ for which
Eqs.~\eqref{eq:m_inf0} and \eqref{eq:m_inf1} admit a positive solution
for $M_{\infty}$. As test beds, we consider neutron stars with central
energy densities which result in a canonical $1.4\,M_{\odot}$ and the
maximum allowed mass in GR, i.e. $2.05\, M_{\odot}$.
As evident from Fig.~\ref{fig:errors}, the error can become significant as we increase
the value of $|\mu_1|$.
By demanding that the errors remain within $5\%$ we can narrow down the admissible values of $\mu_1$ to $[-1.0, 0.1]$.
We observe that while $M_{\infty}$ can deviate greatly from the GR value
(e.g. $M_{\infty}$ reduces by $\approx 21\%$ when $\mu_1 = -1.0$ with respect to a
$1.4 \, M_{\odot}$ neutron star), ${\cal F}$ remains below unity (see right panel of
Fig.~\ref{fig:errors}). This is because large negative values of the
parameter $\mu_1$ make the star less compact (i.e. Newtonian), as
evidenced in Fig.~1 of~\cite{Glampedakis:2015sua}.

We emphasize that the larger errors for some values of $\mu_1$ are not an issue
with the post-TOV formalism itself, but serve to constrain the values of
$\mu_1$ for which the perturbative expansion is valid.
From a practical point of view, excluding large values of $|\mu_1|$ is
a sensible strategy, since the resulting stellar parameters are so
different with respect to their GR values that these cannot be
considered as meaningful post-TOV corrections.  Hereafter, whenever we
refer to $M_{\infty}$ we mean the mass calculated using
Eq.~\eqref{eq:m_inf2} with $\mu_1 \in [-1.0, 0.1]$.

Within this approximation we are free to use the Taylor-expanded form of
(\ref{m_ext}), i.e.
\be
\label{eq:m_inf}
m(r) = M_\infty \left ( 1 - 2\pi \mu_1 M^2_\infty/r^2 \right )\,.
\ee
This expression leads to the exterior $g_{rr}$ metric
\begin{align}
g_{rr} (r) = \left ( 1 - \frac{2 M_\infty}{r} \right )^{-1}  -4\pi \mu_1 \frac{M^3_\infty}{r^3}
+ {\cal O} \left  (\frac{\mu_1 M_\infty^4}{ r^4} \right ). \nonumber \\
\label{g11_ext}
\end{align}
This expression allows us to identify $M_\infty$ as the spacetime's
gravitating mass (see also the result for $g_{tt}$ below).

The next step is to use our result for $m(r)$ in (\ref{dnu_ext1}) and
integrate to obtain $\nu(r)$. After expanding to 2PN post-TOV order we
obtain:
\be
\frac{d\nu}{dr} = \frac{2M_\infty}{r^2} \left ( 1-\frac{2M_\infty}{r} \right )^{-1}
+ 2 \chi \frac{M^3_\infty}{r^4}.
\ee
where the parameter $\chi$, defined in Eq.~\eqref{eq:chi_def},
quantifies the departure from the Schwarzschild metric.
Integrating,
\be
\nu(r) = \log \left ( 1-\frac{2M_\infty}{r} \right ) - \frac{2\chi}{3}\frac{M^3_\infty}{r^3},
\label{eq:nu_sol_ext}
\ee
where the integration constant has been eliminated by requiring asymptotic flatness.
The resulting exterior $g_{tt}$ metric component is
\be
g_{tt} (r) = -\left ( 1-\frac{2M_\infty}{r} \right )  + \frac{2\chi}{3} \frac{M^3_\infty}{r^3}
+{\cal O}  \left ( \frac{\chi M_\infty^4}{r^4} \right  )\,.
\label{eq:g00_ext}
\ee

Eqs.~(\ref{eq:g00_ext}) and (\ref{g11_ext}) represent our final
results for the 2PN-accurate exterior post-Schwarzschild metric.  From
this construction it follows that post-TOV stars for which
$\mu_1 = \mu_2 = \pi_2 = 0$ have the Schwarzschild metric as the
exterior spacetime. The following sections describe how the exterior
metric  can be used to compute observables of relevance for
neutron star astrophysics.


\section{Surface redshift  \& stellar compactness}
\label{sec:redshift}

\begin{figure}[htb]
\includegraphics[width=0.45\textwidth]{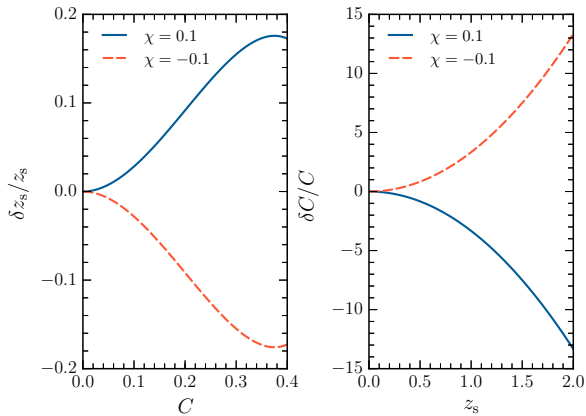}
\caption{{\it Surface redshift and stellar compactness.} Relative percent changes
with respect to GR for both $z_{\rm s}$ and $C$ for different values of $\chi$.
}
\label{fig:redshift}
\end{figure}

The surface redshift is among the most basic neutron star observables
that could be affected by a change in the gravity theory. The surface
redshift is defined in the usual way as
\be
z_s \equiv \frac{\lambda_{\infty} - \lambda_{\rm s}}{\lambda_{\rm s}}
= \frac{f_{\rm s}}{f_{\infty}} - 1,
\label{eq:defredshift}
\ee
where $\lambda$ and $f$ are the wavelength and frequency of a photon,
respectively. Here and below, the subscripts $s$ and $\infty$ will
denote the value of various quantities at the stellar surface $r = R$
and at spatial infinity. The familiar redshift formula
\begin{equation}
\frac{f_\infty}{f_{\rm s}} = \left[\frac{g_{tt} (R)}{g_{tt} (\infty)}\right]
^{1/2}
\end{equation}
is valid for any static spacetime, regardless of the form of the field equations.
Using the metric (\ref{eq:g00_ext}), we easily obtain (at first post-TOV order)
\be
\frac{f_{\rm s}}{f_{\infty}}  = \left( 1 - \frac{2 M_{\infty}}{R}\right)^{-1/2} + \frac{\chi}{3} \frac{M_{\infty}^3}{R^3}.
\ee
Given that the frequency shift depends only on the ratio $M_\infty/R$, it is more convenient to work in terms of the
stellar compactness
\be
C = {M_\infty}/{R}.
\ee
Then from the definition of the surface redshift we obtain
\be
z_{\rm s}  = z_{\rm GR} + \frac{\chi}{3} C^3,
\label{eq:ptovredshift}
\ee
where
\be
z_{\rm GR} \equiv \left (\, 1 - 2C \, \right)^{-1/2} - 1,
\ee
is the standard redshift formula in GR, while the second term represents the
post-TOV correction. Observe that $z_{\rm s}$ can be smaller or larger
than $z_{\rm GR}$ depending on the sign of the parameter $\chi$.
This is shown in Fig.~\ref{fig:redshift} (left panel), where we plot the
percent difference $\delta z_{\rm s}/z_{\rm s}
\equiv 100 \times (z_s - z_{\rm GR})/z_{\rm GR} $
as a function of $C$ for two representative cases ($\chi = \pm 0.1$).

A characteristic property of the redshift is that it is a function of
$C$, and as such it cannot be used to disentangle mass and radius
individually. A given \textit{observed} surface redshift $z_{\rm obs}$
can be experimentally interpreted either as
$z_{\rm obs} = z_{\rm GR} (C)$ or $z_{\rm obs} =z_s (C,\chi)$, and
therefore lead to different estimates for $C$ (for a given $\chi$).
Fig.~\ref{fig:redshift} (right panel), where we plot the percent
difference $\delta C / C \equiv 100 \times (C-C_{\rm GR})/C_{\rm GR}$ as a
function of $z_{\rm s}$, shows how much the inferred compactness would differ
in the two cases where $\chi = \pm 0.1$. A positive (negative) $\chi$ leads to a
lower (higher) inferred compactness with respect to GR. The figure
suggests that the ``error'' in $C$ becomes significant for redshifts
$z_{\rm s} \gtrsim 1$.

It is a straightforward exercise to invert the redshift formula and obtain
a post-TOV expression $C= C(z_{\rm s})$. We first write
\be
1 + z_{\rm s} = \frac{1}{\sqrt{-g_{tt} (R)}} \,\Rightarrow\,
g_{tt} (R) = -\frac{1}{(1+z_{\rm s})^2}.
\ee
Upon inserting the post-Schwarzschild metric (\ref{eq:g00_ext}) we get a
cubic equation for the compactness,
\be
-1  + \frac{1}{(1+z_{\rm s})^2} + 2 C + \frac{2}{3} \chi C^3 = 0.
\ee
In solving this equation we take into account that the small parameters
are $C$ and $z_{\rm s} \sim {\cal O}(C)$ and that the $\chi \to 0$ limit should be smooth. We find,
\be
C = C_{\rm GR} \left (\, 1   - \frac{1}{3} \chi z_{\rm s}^2   \, \right  ),
\label{Cptov}
\ee
where
\be
C_{\rm GR} = \frac{1}{2} \left [ 1 - (1+z_{\rm s})^{-2} \right ],
\ee
is the corresponding solution in GR.

As was the case for the post-TOV redshift formula, the compactness of a post-TOV star can be pushed
above (below) the GR value by choosing a negative (positive) parameter $\chi$.

The two main results of this section, Eqs.~(\ref{eq:ptovredshift}) and
(\ref{Cptov}), are also interesting from a different prespective,
namely, their dependence on the single post-TOV parameter $\chi$. This
dependence entails a degeneracy with respect to the coefficient triad
$\{ \mu_1, \mu_2, \pi_2 \}$ when (for example) a neutron star redshift
observation is used as a gravity theory discriminator.  The
redshift/compactness $\chi$-degeneracy is another reminder of the
intrinsic difficulty in distinguishing non-GR theories of gravity from
neutron star physics (see e.g. discussion around Fig. 1
in~\cite{Glampedakis:2015sua}).

\section{Bursting neutron stars}
\label{sec:burst}

A potential testbed for measuring deviations from GR with a
parametrized scheme like our post-TOV formalism is provided by
accreting neutron stars exhibiting the so-called type I bursts. These
are X-ray flashes powered by the nuclear detonation of accreted matter
on the stellar surface layers~\cite{1993SSRv...62..223L}. The luminosity associated
with these events can reach the Eddington limit and may cause a photospheric
radius expansion (see e.g.~\cite{Kuulkers:2002db,Steiner:2010fz}), thus offering a number of
observational ``handles" to the system (see below for more details).

A paper by Psaltis~\cite{Psaltis:2007rv} proposed type I bursting
neutron stars as a means to constrain possible deviations
from GR.  Psaltis' analysis, based on a static and spherically
symmetric model for describing the spacetime outside a non-rotating
neutron star, is general enough to allow a direct adaptation to the
post-TOV scheme. For that reason we can omit most of the technical
details discussed in~\cite{Psaltis:2007rv} and instead focus on the
key results derived in that paper.

There is a number of observable quantities associated with type I
bursting neutron stars that can be used to set up a test of GR.  The
first one is the surface redshift $z_{\rm s}$; in
Section~\ref{sec:redshift} we have derived post-TOV formulae for $z_{\rm s}$
and the stellar compactness $C$, which are used below in the
derivation of a constraint equation between the post-Schwarzschild
metric and the various observables.

The luminosity (as measured at infinity) of a source located at a
(luminosity) distance $D$ is
\be
L_\infty = 4\pi D^2 F_\infty,
\ee
where $F_\infty$ is the (observable) flux. This luminosity can be written in a black-body form with the
help of an apparent surface area $S_{\rm app}$ and a color temperature (as measured at infinity) $\bar{T}_{\infty}$:
\be
4\pi D^2 F_\infty = \sigma_{\rm SB} S_{\rm app} \bar{T}^4_\infty,
\ee
where $\sigma_{\rm SB}$ is the Stefan-Boltzmann constant. We then
define the second observable parameter used in this analysis, i.e. the
apparent radius
\be
R_{\rm app} \equiv \left ( \frac{S_{\rm app}}{4\pi} \right )^{1/2} = D \left ( \frac{F_\infty}{\sigma_{\rm SB} \bar{T}_\infty^4} \right )^{1/2}.
\label{Rapp1}
\ee
As evident from its form, $R_{\rm app}$ is independent of the
underpinning gravitational theory, at least to the extent that the
theory does not appreciably modify the (luminosity) distance to the
source.

The surface color temperature is related to the intrinsic effective temperature $T_{\rm eff}$ via the standard
color correction factor $f_{\rm c}$~\cite{Majczyna:2004bs,Suleimanov:2010bp},
\be
\bar{T}_{\rm s} = f_{\rm c} T_{\rm eff}.
\ee
 The observed temperature at infinity picks up a redshift factor with respect to its local surface value, that is,
\be
\bar{T}_\infty = f_{\rm c} \sqrt{-g_{tt} (R)} T_{\rm eff}.
\ee
The effective temperature is the one related to the source's intrinsic luminosity,
\be
L_{\rm s} = 4\pi R^2 \sigma_{\rm SB} T_{\rm eff}^4.
\ee
As shown in~\cite{Psaltis:2007rv},
\be
L_\infty = - g_{tt} (R) L_{\rm s} = 4\pi R^2 \sigma_{\rm SB} \left ( \frac{\bar{T}_\infty}{f_{\rm c}} \right )^4 [-g_{tt} (R)]^{-1}.
\ee
Combining this with the preceding formulae leads to,
\be
\frac{R_{\rm app}}{R} = \frac{1+z_{\rm s}}{f_{\rm c}^2}.
\label{Rapp2}
\ee
The third relevant observable is the Eddington luminosity/flux at infinity. This is given by~\cite{Psaltis:2007rv},
\be
L_{\rm E}^\infty =  4\pi D^2 F^\infty_{\rm E} =  \frac{4\pi}{\kappa} \frac{R^2}{(1+z_{\rm s})^2} g_{\rm eff},
\label{Ledd1}
\ee
where $\kappa$ is the opacity of the matter interacting with the radiation field\footnote{Typically, this interaction
manifests itself as Thomson scattering in a hydrogen-helium plasma, in which case the opacity is
$\kappa \approx 0.2 (1+ X)\,\mbox{cm}^2/\mbox{gr}$ where $X$ is the hydrogen mass fraction~\cite{Steiner:2010fz}.}
and $g_{\rm eff}$ is an effective surface gravitational acceleration, defined as:
\be
g_{\rm eff} =\frac{1}{2\sqrt{g_{rr} (R)}} \frac{g^\prime_{tt} (R)}{g_{tt} (R)}.
\ee
This parameter is key to the present analysis as it encodes the departure from the general relativistic Schwarzschild metric.

\begin{figure}[thb]
\includegraphics[width=0.44\textwidth]{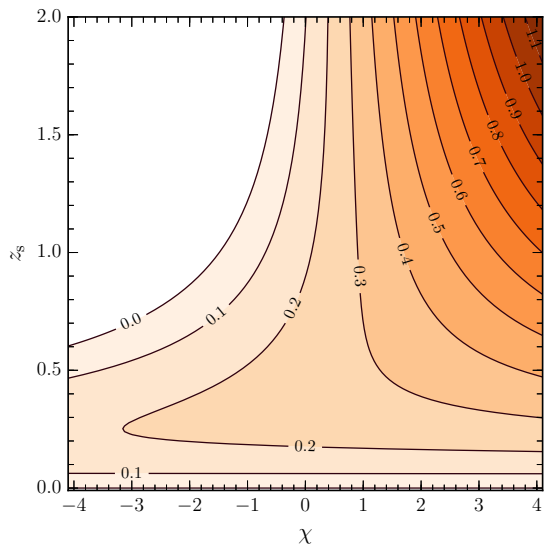}
\caption{{\it Bursting neutron star constraints.}  The surfaces are
  contours of constant
  $ \left [\, 1 + (2/3) \chi z_{\rm s}^2 \, \right ] z_{\rm
    s}(2+z_{\rm s})/(1+z_{\rm s})^4 $
  in the $(z_{\rm s},\,\chi)$ plane.  This quantity is a combination
  of observables -- cf. the right-hand side of Eq.~(\ref{constraint})
  -- and therefore it is potentially measurable; a measurement will
  single out a specific contour in this plot. A further measurement of
  (say) the redshift $z_{\rm s}$ corresponds to the intersection
  between one such contour and a line with $z_{\rm s}={\rm const.}$,
  so it can lead to a determination of $\chi$.}
\label{fig:constraint}
\end{figure}

Having at our disposal the above three observable combinations, the
strategy is to combine them and derive a constraint equation between
the observables and the spacetime metric. To this end, we first need
to eliminate the not directly observable stellar radius $R$ between
(\ref{Rapp2}) and (\ref{Ledd1}) and subsequently solve with respect to
$g_{\rm eff}$. We obtain
\be
g_{\rm eff} =  \kappa \sigma_{\rm SB} \frac{F_{\rm E}^\infty}{F_\infty} \left ( \frac{\bar{T}_\infty}{f_{\rm c}} \right )^4 (1+z_{\rm s})^4.
\label{geff1}
\ee
The remaining task is to express $ g_{\rm eff} $  in terms of $z_{\rm s}$. Using the post-Schwarzschild
metric, Eqs.~ (\ref{g11_ext}) and (\ref{eq:g00_ext}), in  (\ref{geff1}) we obtain the post-TOV result
\be
g_{\rm eff} = \frac{C}{R} (1+z_{\rm s} ) \left (\, 1 + \chi C^2  \, \right ).
\ee
Making use of Eq.~(\ref{Cptov}) for the compactness leads to the desired result [cf. Eq.~(39) in \cite{Psaltis:2007rv}]:
\be
g_{\rm eff} = \frac{z_{\rm s}}{2R} \frac{(2+z_{\rm s})}{(1+z_{\rm s})} \left (\, 1 + \frac{2}{3} \chi z_{\rm s}^2  \, \right ),
\label{geffptov}
\ee
where, as evident, the prefactor represents the GR result.  Finally, after eliminating $R$ with the help
of (\ref{Rapp2}) and (\ref{Rapp1}), we obtain the ``observable" effective gravity:
\be
g_{\rm eff} =  \frac{z_{\rm s}(2+z_{\rm s})}{2 D f_{\rm c}^2}   \left (\, 1 +  \frac{2}{3} \chi z_{\rm s}^2  \, \right )
\left ( \frac{\sigma_{\rm SB} \bar{T}^4_\infty}{F_\infty} \right )^{1/2}.
\label{geff_obs}
\ee
This can then be combined with (\ref{geff1}) to give
\be
 \frac{z_{\rm s}(2+z_{\rm s})}{(1+z_{\rm s})^4}   \left (\, 1 + \frac{2}{3} \chi z_{\rm s}^2   \, \right )
 = 2D \kappa \frac{F_{\rm E}^\infty }{ f_{\rm c}^2} \left ( \frac{\sigma_{\rm SB} \bar{T}_\infty^4 }{F_\infty} \right )^{1/2},
 \label{constraint}
\ee
and consequently
\be
\chi = \frac{3}{2 z^2_{\rm s}} \left [\, 2D \kappa   \frac{ (1+z_{\rm s})^4}{z_{\rm s}(2+z_{\rm s})}  \frac{F_{\rm E}^\infty }{ f_{\rm c}^2}
\left ( \frac{\sigma_{\rm SB} \bar{T}_\infty^4 }{F_\infty} \right )^{1/2}  - 1 \, \right ].
\label{chi_obs}
\ee
This equation is the main result of this section and provides, at
least as a proof of principle, a quantitative connection between the
post-Schwarzschild correction to the exterior metric [in the form of
the $\chi$ coefficient defined in Eq.~\eqref{eq:chi_def}] and
observable quantities in a type I bursting neutron
star. Ref.~\cite{Psaltis:2007rv} arrives at a similar result [their
Eq.~(49)] which has the same physical meaning, but is not identical to
Eq.~(\ref{constraint}) due to the different assumed form of the
exterior metric.

Our results are illustrated in Fig.~\ref{fig:constraint}, where we
show the left-hand side of Eq.~(\ref{constraint}) as a contour plot in
the $(z_{\rm s},\,\chi)$ plane. Each contour represents a specific
measurement of this observable quantity. An additional surface
redshift measurement can lead, at least in principle, to the
determination of the post-TOV parameter $\chi$, as given by
Eq.~(\ref{chi_obs}).


\section{Quasi-periodic oscillations}
\label{sec:qpos}

The post-Schwarzschild metric allows us to compute the geodesic motion
of particles in the exterior spacetime of post-TOV neutron stars.
Geodesics in neutron star spacetimes play a key role in the
theoretical modelling of the QPOs observed in the X-ray spectra of
accreting neutron stars.
The detailed physical mechanism(s) responsible for the QPO-like time
variability in the flux of these systems is still a matter of debate,
but some of the most popular models are based on the notion of a
radiating hot ``blob'' of matter moving in nearly circular geodesic
orbits. The QPO frequencies are identified either with the orbital
frequencies, or with simple combinations of the orbital
frequencies. The most popular models are variants of the relativistic
precession~\cite{StellaApJ98,StellaPRL99} and epicyclic
resonance~\cite{Abramowicz:2001bi} models.

In this section we discuss the relevant orbital frequencies within  the post-TOV
formalism and derive formulae that could easily be  used in the aforementioned QPO
models. In principle, matching the orbital frequencies to the QPO data would allow
one to extract post-TOV parameters such as $\chi$ and $\mu_1$
(see~\cite{Ryan95,PappasMNRAS2012,PappasMNRAS2015ST} for a similar exercise in the
context of GR and scalar-tensor theory).

For nearly circular orbits in a spherically symmetric spacetime, the only
perturbations of interest are the radial ones (i.e., there is periastron precession
but no Lense-Thirring nodal precession) and therefore we can associate two
frequencies to every circular orbit: the orbital azimuthal frequency of the
circular orbit $\Omega_\varphi$ and the radial epicyclic frequency $\Omega_r$.

Geodesics in a static, spherically symmetric spacetime are
characterized by the two usual conserved quantities, the energy
$E =-g_{tt} \dot{t}$ and the angular momentum
$ L= g_{\phi\phi} \dot{\phi}$. Here both constants are defined per
unit particle mass, and the dots stand for differentiation with
respect to proper time. The four-velocity normalization condition
$u^au_a=-1$ yields an effective potential equation for the particle's
radial motion,
\be
g_{rr} \dot{r}^2 = -\frac{E^2}{g_{tt}}-\frac{L^2}{g_{\phi\phi}}-1
\equiv V_{\rm eff}(r).
\label{geodesic}
\ee
The conditions for circular orbits are
$V_{\rm eff}(r)=V'_{\rm eff}(r)=0$, where the prime denotes
differentiation with respect to the radial coordinate.  Hereafter $r$
will denote the circular orbit radius. From these conditions we can
determine the orbital frequency
$\Omega_\varphi \equiv\dot{\phi}/\dot{t}$ measured by an observer at
infinity.\footnote{Apart from its implications for the QPOs, the
  post-TOV corrected orbital frequency would imply a shift in the
  corotation radius $r_{\rm co}$ in an accreting system. This radius
  is defined as $\Omega_*= \Omega_\varphi (r_{\rm co})$, where
  $\Omega_*$ is the stellar angular frequency, and plays a key role in
  determining the torque-spin equilibrium in magnetic field-disk
  coupling models. Using the above definition we find the following result for
  the post-TOV corotation radius:
  $r_{\rm co}= M_{\infty}(M_{\infty} \Omega_*)^{-2/3} \left [ \, 1 + (\chi/3) (M_{\infty} \Omega_*)^{4/3}  \, \right ]$.  }
  The square of the orbital frequency is then given as
\begin{align}
\Omega_\varphi^2 &= -\frac{g'_{tt}}{g'_{\phi\phi}} =
\frac{M_{\infty}}{r^3}\left(1+\chi\frac{M_{\infty}^2}{r^2}\right).
\label{orbital}
\end{align}
The Schwarzschild frequency is recovered for
$\chi=0$.

\begin{figure}[htb]
\includegraphics[width=0.44\textwidth]{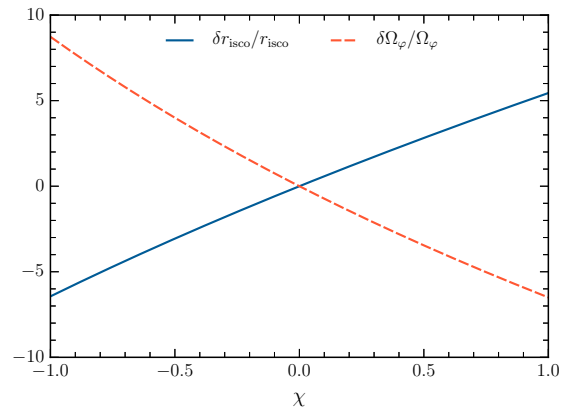}
\caption{{\it ISCO quantities.}  ISCO quantities as functions of
  $\chi$. The solid curve corresponds to the relative difference (in
  percent) of $r_{\rm isco}$ with respect to GR, while the dashed
  curve corresponds to the relative difference of the orbital
  frequency at the ISCO, $ ( \Omega_\varphi )_{\rm isco}$.}
\label{fig:ISCO}
\end{figure}

The radial epicyclic frequency can be calculated from the
equation for radially perturbed circular orbits, which
follows from Eq.~(\ref{geodesic}):
\begin{subequations}
\begin{align}
\Omega_r^2 &= -\frac{g^{rr}}{2 \dot{t}^2} V''_{\rm eff}(r)\approx \frac{M_{\infty}}{r^3}
\left[1-\frac{6 M_{\infty}}{r} -\frac{\chi  M_{\infty}^2}{r^2}  + {\cal O} (r^{-3})\right] \label{epicyclic1}\\
&=  \Omega_\varphi^2 \left[1-\frac{6 M_{\infty}}{r} -\frac{2 \chi  M_{\infty}^2}{r^2}  +  {\cal O} (r^{-3}) \right], \label{epicyclic3}
\end{align}
\end{subequations}
where again the frequency is calculated with respect to observers at
infinity.
From the post-TOV expanded result we can see that the first two terms correspond to the Schwarzschild
epicyclic frequency.
The additional post-TOV terms in these formulae produce a shift in the
frequency and radius of the innermost stable circular orbit (ISCO)
with respect to their GR values -- the latter quantity is determined
by the condition $\Omega_r^2=0$, which in GR leads to the well-known
result $r_{\rm isco}=6M_{\infty}$. The corresponding post-TOV ISCO is obtained
from (\ref{epicyclic1}), up to linear order in $\chi$, as:
\be
r_{\rm isco} \approx 6 M_{\infty} \left(1+\frac{19}{324}\chi \right ).
\label{risco}
\ee
The post-TOV ISCO parameters $r_{\rm isco}$ and $(\Omega_\varphi)_{\rm isco}$
are plotted in Fig.~\ref{fig:ISCO} as functions of the parameter $\chi$. As evident
from Eq.~(\ref{risco}), a positive (negative) $\chi$ implies
$r_{\rm isco} > 6M_{\infty}$ ($r_{\rm isco} < 6 M_{\infty}$).
If one takes Eq.~(\ref{epicyclic1}) at face value for the given
post-Schwarzschild metric, for negative enough values of $\chi$
there is no ISCO solution, but this occurs well beyond the point where
it is safe to use our perturbative formalism.
The orbital frequency profile remains rather simple, with
$(\Omega_\varphi)_{\rm
  isco}$ exceeding the GR value when $r_{\rm isco} < 6
M_{\infty}$ (and vice versa).

Besides the frequency pair $\{\Omega_\varphi, \Omega_r \}$, a third prominent
quantity in the QPO models is the frequency
\be
\Omega_{\rm per}=\Omega_\varphi-\Omega_r,
\ee
associated with the orbital periastron precession (for example, in the
relativistic precession model~\cite{StellaApJ98,StellaPRL99} this
frequency is typically associated with the low-frequency QPO) .

Given our earlier results, it is straightforward to derive a series
expansion in powers of $1/r$ for $\Omega_{\rm per}$.
However, it is usually more desirable to produce a series expansion with respect
to an observable quantity, such as the circular orbital velocity
$U_{\infty}=(M_{\infty}\Omega_\varphi)^{1/3}$. This can be done by first expanding
$U_\infty$ with respect to $1/r$ and then inverting the expansion, thus
producing a series in $U_\infty$. The outcome of this recipe is
\begin{align}
 \frac{\Omega_{\rm per}}{\Omega_\varphi} &= 1-\frac{\Omega_r}{\Omega_\varphi}
 =   3 U_{\infty}^2  +\left(\frac{9}{2} + \chi\right) U_{\infty}^4 +    {\cal O} (U_{\infty}^{6}).
 \label{periastron1}
\end{align}
A similar ``Keplerian" version of this expression can be produced if
we opt for using the velocity $U_{\rm K}$ and mass $M_{\rm K}$ that an
observer would infer from the motion of (say) a binary system under
the assumption of exactly Keplerian orbits. These are
$U_{\rm K} =(M_{\rm K}\Omega_\varphi)^{1/3}$ and
$M_{\rm K}=r^3\Omega^2_\varphi$, so that
$M_{\rm K} =M_{\infty}\left(1+\chi M_{\infty}^2/r^2\right)$.  The
resulting series is identical to Eq.~(\ref{periastron1}) when
truncated to $U_{\rm K}^4$ order.  Higher-order terms, however, are
different (see the following section).

The above results for the frequencies
$\{\Omega_\varphi, \Omega_r, \Omega_{\rm per}\}$ suggest that a
QPO-based test of GR within the post-TOV formalism could in principle
allow the extraction of the post-TOV parameter $\chi$.  In this sense
these frequencies probe the same kind of deviation from GR (and suffer
from the same degree of degeneracy) as the observations of bursting
neutron stars discussed in Section~\ref{sec:burst}.

\begin{figure}[thb]
\includegraphics[width=0.44\textwidth]{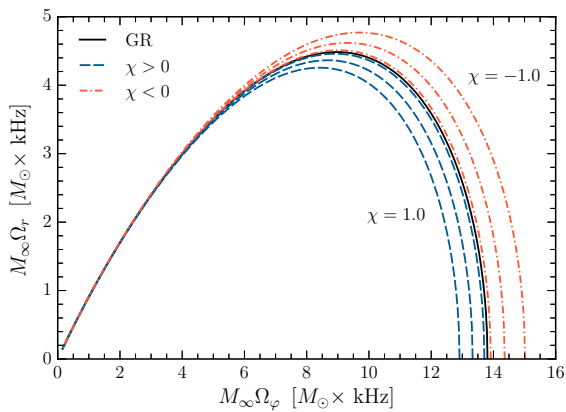}
\caption{{\it Orbital frequencies.}
Plots of $\Omega_r$ against $\Omega_{\varphi}$ for
different values of $\chi=\pm0.1,\, \pm 0.5$ and $\pm1$. The black solid curve corresponds to the GR case.
The dashed curves correspond to positive values of $\chi$ (and $r_{\rm isco}>6M_{\infty}$)
while the dashed-dotted curves correspond to negative values of $\chi$ (and
$r_{\rm isco}<6M_{\infty}$).
}
\label{fig:freq}
\end{figure}

We conclude this section by sketching how this procedure works in
practice in the context of the relativistic precession model. The twin
$\mbox{kHz}$ QPO frequencies $\{\nu_1, \nu_2 \}$ seen in the flux of
bright low-mass X-ray binaries are identified with the azimuthal and
periastron precession orbital frequencies. More specifically, the
high-frequency member of the pair is identified with the azimuthal
frequency ($\nu_2 = \nu_\varphi = \Omega_\varphi/2\pi$), while the
low-frequency member is identified with the periastron precession
($\nu_1 = \nu_{\rm per} = \Omega_{\rm per}/2\pi$). With this
interpretation, the QPO separation is equal to the radial epicyclic
frequency: $\Delta \nu = \nu_2 - \nu_1 = \Omega_r/2\pi$.

We use our previous results [Eqs.~(\ref{orbital}), (\ref{epicyclic3}),
(\ref{periastron1})] to plot these orbital frequencies (clearly,
$\nu_1/\nu_2 = \Omega_{\rm per}/\Omega_\varphi$ and
$\Delta \nu/\nu_2 = \Omega_r/\Omega_\varphi $) as functions of each
other and for varying $\chi$. As it turns out, deviations from GR are
best illustrated by plotting $\Omega_r (\Omega_\varphi)$ (or
equivalenty $\Delta \nu (\nu_2)$). In Fig.~\ref{fig:freq} we plot the
dimensionless combinations
$M_\infty \Omega_r,~M_\infty \Omega_\varphi$ (in units of kHz for the frequencies and solar masses for $M_{\infty}$). As
we can see, the post-TOV models considered here ($ -1 <\chi < 1 $) are
qualitatively similar to the GR result (black solid curve), all cases
showing the characteristic hump in $\Omega_r$ as $\Omega_\varphi$
increases (so that the orbital radius decreases).  This feature is
evidently associated with the existence of an ISCO (where
$\Omega_r \to 0$) and is consistent with a similar trend seen
in observations~\cite{StellaPRL99}.


\section{Multipolar structure of the spacetime}
\label{sec:multipoles}

Expansions like Eq.~(\ref{periastron1}) contain information about the
multipolar structure of the background spacetime. That expansion can
be directly compared against a similar expansion derived by
Ryan~\cite{Ryan95} for an axisymmetric, stationary spacetime in GR
with an arbitrary set of mass ($M_0=M_{\infty}, M_2, M_4, ...$) and current
($S_1, S_3, S_5, ...$) Geroch-Hansen multipole
moments~\cite{Geroch:1970cc,Geroch:1970cd,Hansen:1974zz}:\footnote{A
  multipolar expansion in scalar-tensor theory can be found
  in~\cite{Pappas:2015moments}.  Specific calculations were also
  carried out in other theories: for example, the quadrupole moment
  was computed in Einstein-dilaton-Gauss-Bonnet gravity
  \cite{Kleihaus:2014quadrupole}.}
\begin{align}
\frac{\Omega_{\rm per}}{\Omega_\varphi} &= 3 U^2 - 4\frac{S_1}{M_{\infty}^2} U^3
+ \left (\, \frac{9}{2} - \frac{3}{2} \frac{M_2}{M_{\infty}^3} \,\right ) U^4 -10 \frac{S_1}{M_{\infty}^2} U^5
\nonumber \\
& + \left (\, \frac{27}{2} -2\frac{S_1^2}{M_{\infty}^4} -\frac{21}{2} \frac{M_2}{M_{\infty}^3}  \, \right ) U^6 +  {\cal O} (U^{7} ).
\label{ryan}
\end{align}
where $U = (M_{\infty} \Omega_\varphi )^{1/3} $ denotes the orbital velocity.

To understand the PN accuracy of the post-TOV expansion in this context, it is useful to consider the
effect of higher PN order terms in the metric. Imagine that the $g_{tt}$ and $g_{rr}$ metric components
[see Eqs.~(\ref{g11_ext}), (\ref{eq:g00_ext})] included 3PN corrections of the schematic form,
\begin{align}
g_{tt}(r)&=g_{tt}^{\rm 2PN}+\alpha_{tt} \frac{M_{\infty}^3}{r^3},
\label{eq:gtt3PN}\\
g_{rr}(r)&=g_{rr}^{\rm 2PN}+\alpha_{rr} \frac{M_{\infty}^3}{r^3}.
\end{align}

We can use the coefficients $\alpha_{tt}$ and $\alpha_{rr}$ as bookeeping parameters in order to
understand how these omitted higher-order contributions affect the results of the previous section.
The recalculation of the various expressions reveals that the orbital frequency remains
unchanged to 2PN order; the 3PN term of Eq.~(\ref{eq:gtt3PN}) contributes at the next
order, as expected. The same applies to the epicyclic frequency, as we
can see for example from the modified Eq.~(\ref{epicyclic3}), where
the next-order correction is a mixture of $g_{rr}^{\rm 2PN}$  and the 3PN term in $g_{tt}$:
\begin{align}
\Omega_r^2&=\Omega_\varphi^2 \left[1-\frac{6 M_{\infty}}{r} -\frac{2 \chi  M_{\infty}^2}{r^2} \right.
\nonumber \\
&\left.
+ \frac{(4 \pi  \mu_1 -6\alpha_{tt})M_{\infty}^3}{r^3} +  {\cal O} (r^{-4}) \right].
\end{align}
Proceeding in a similar way we find the next-order correction to the
Ryan-like expansion (\ref{periastron1}):
\begin{align}
 \frac{\Omega_{\rm per}}{\Omega_\varphi} &=   3 U_{\infty}^2  +\left(\frac{9}{2} + \chi\right) U_{\infty}^4 
 \nonumber \\
 &+\left[\frac{27}{2} + 2 (\chi -\pi  \mu_1 )+3\alpha_{tt}\right] U_{\infty}^6
+   {\cal O} (U_{\infty}^{8}).
\label{periastron2}
\end{align}
We can see that the 3PN term ``contaminates"  the PN correction that was omitted in
Eq.~(\ref{periastron1}). Repeating the same exercise for the Keplerian version of the
multipole expansion (i.e. where the orbital velocity $U_\infty$ is replaced by $U_{\rm K}$) we find:
\begin{align}
 \frac{\Omega_{\rm per}}{\Omega_\varphi} &= 3 U^2_{\rm K}  +\left(\frac{9}{2} + \chi \right) U^4_{\rm K} \nonumber \\
 &+\left(\frac{27}{2} - 2 \pi \mu_1+3\alpha_{tt}\right) U^6_{\rm K}  
+   {\cal O} (U^{8}_{\rm K} ).
\label{periastron3}
\end{align}
At the PN order considered in the previous section the two expressions were identical 
but, as we can see, they differ at the next order.

We now have Ryan-type multipole expansions of the post-Schwarzschild
spacetime up to 3PN in the circular orbital velocity, which we can
compare against Eq.~(\ref{ryan}) to draw (with some caution) analogies
and differences between GR and modified theories of gravity.

For instance, odd powers of $U_\infty$ are missing in
Eq.~(\ref{periastron2}) because the nonrotating post-Schwarzschild
spacetime has vanishing current multipole moments.
Furthermore, we can see that the quadrupole moment $M_2$, first
appearing in the coefficient of $U^4_\infty$ in Eq.~(\ref{ryan}), can
be associated with $\chi$. The parameter $\chi$ is an
\textit{effective} quadrupole moment in the sense that
\be
M_2^{\rm eff} = -\frac{2}{3} \chi M_{\infty}^3.
\label{M2eff}
\ee
Indeed, this relation implies that a positive (negative) $\chi$ could
be associated with an oblate (prolate) source of the gravitational
field.

The identification (\ref{M2eff}) holds at
$ {\cal O} (U_{\infty}^{4})$. The next-order term $U^6_\infty$ would,
in general, lead to a different effective $M_2$. Hence, the comparison
between the $U_{\infty}^4$ and $U_{\infty}^6$ terms could provide a
null test for the GR-predicted quadrupole. However, there is a special
case where these two terms could be consistent with the same effective
quadrupole (\ref{M2eff}): this occurs when the post-TOV parameters
satisfy the condition $5\chi = -2\pi\mu_1+3\alpha_{tt}$, in which case
the expansion (\ref{periastron2}) behaves as a ``GR mimicker".

A different kind of ``multipole'' expansion in powers of $1/r$ can be
applied to the metric functions $\nu(r), m(r)$ [see
Eqs. (\ref{dnu_ext0})--(\ref{eq:nu_ext_gr})], leading to an alternative
calculation of the ADM mass $M_\infty$ of a post-TOV star. We consider
the expansions
\be
\nu(r) = \sum_{n = 0}^{\infty} \frac{\nu_{n}}{r^n}, \qquad
m(r) = \sum_{n = 0}^{\infty} \frac{m_{n}}{r^n},
\label{eq:expansion}
\ee
where $\nu_n$ and $m_n$ are constant coefficients. In addition,
we impose that $\nu_0 = 0$ and $m_0 = M_{\infty}$. We subsequently substitute
these expansions into Eqs. (\ref{dnu_ext0})--(\ref{dm_ext1}), expand for
$r/R \gg 1$, and then solve for the coefficients. The outcome of this exercise
in the vacuum exterior spacetime is
\begin{align}
m(r) &= M_{\infty} - 2 \pi \mu_1 \frac{M_{\infty}^3}{r^2} + {\cal O}(r^{-4})\,,
\label{eq:m_exp}
\\
\nonumber \\
\nu(r) &= - \frac{2 M_{\infty}}{r} - \frac{2 M_{\infty}^2}{r^2}
 - \frac{2}{3} \frac{M_{\infty}^3}{r^3}\left( 4 + \chi \right) + {\cal O}(r^{-4})\,.
\label{eq:nu_exp}
\end{align}
As expected, the top equation is consistent with our earlier result,
Eq.~(\ref{eq:m_inf}). To get an agreement between
Eqs.~(\ref{eq:nu_sol_ext}) and (\ref{eq:nu_exp}) we must expand the
logarithm appearing in the former equation in powers of
$M_{\infty}/r$, thus recovering Eq.~(\ref{eq:nu_exp}).


\section{Conclusions}
\label{sec:conclusions}

In this paper we have demonstrated the applicability of the post-TOV
formalism to a number of facets of neutron star astrophysics. Let us
summarize our main results.
The exterior post-Schwarzschild metric [Eqs.~(\ref{g11_ext}) and
(\ref{eq:g00_ext})] depends only on the ADM mass $M_\infty$ (given by
Eq.~(\ref{eq:m_inf2})) and on just two post-TOV parameters $\mu_1$ and
$\chi$. These are subsequently used to produce a post-TOV formula for
the surface redshift, Eq.~(\ref{eq:ptovredshift}), which is a function
of the stellar compactness and $\chi$.  Next, we have shown how a
basic post-TOV model for type I bursting neutron stars can be
constructed.  The key equation here is (\ref{chi_obs}), which gives
$\chi$ (the only post-TOV parameter appearing in the model) in terms
of observable quantities. We also computed geodesic motion in the
post-Schwarzschild exterior of post-TOV neutron star models, finding
expressions for the orbital, epicyclic and periastron precession
frequencies of nearly circular orbits [Eqs.~(\ref{orbital}),
(\ref{epicyclic3}), (\ref{periastron1})] and for the ISCO radius
[Eq.~(\ref{risco})]. These results can be fed into models for QPOs
from accreting neutron stars, such as the relativistic precession
model.  Finally, on a more theoretical level, we have sketched how the
post-TOV parameters enter in the spacetime's multipolar structure
[Eq.~(\ref{periastron3})].

The meticulous reader may have noticed that, in spite of the exterior
metric being a function $g_{tt} (\chi)$ and $g_{rr} (\mu_1)$, all
other post-TOV results feature only $\chi$, while $\mu_1$ is either
not present or enters at higher order. This is not a coincidence:
these quantities either depend solely on $g_{tt}$ (e.g. the redshift)
or receive their leading-order contributions from $g_{tt}$ (e.g. the
orbital frequencies).

The post-TOV formalism developed in~\cite{Glampedakis:2015sua} and in
this paper can be viewed as a basic ``stage-one'' version of a more
general framework. There are several directions one can follow for
taking the formalism to a more sophisticated level, and here we
discuss just a couple of possibilities.

An obvious improvement is the addition of stellar rotation. This is
necessary because all astrophysical compact stars rotate, some of them
quite rapidly, and the influence of rotation is ubiquitous, affecting
to some extent all of the effects discussed in this paper. As a first
stab at the problem, it would make sense to work in the Hartle-Thorne
slow rotation approximation~\cite{Hartle:1967he,Hartle:1968si}, which
should be accurate enough for all but the fastest spinning neutron
stars~\cite{Berti:2004ny}.

There are equally important possibilities for improvement on the
modified gravity sector of the formalism. The present post-TOV theory
is oblivious to the existence of dimensionful coupling constants, such
as the ones appearing in many modified theories of gravity (e.g.
$f(R)$ theories or theories quadratic in the curvature). These
coupling parameters should be added to the existing set of fluid
parameters, and participate in the algorithmic generation of
``families" of post-TOV terms (see~\cite{Glampedakis:2015sua} for
details). The extended set of parameters will most likely lead to a
proliferation of post-TOV terms, and result in more complicated
stellar structure equations than the ones used so far
[i.e. Eqs.~(\ref{eq:PTOV_2PN_intro})]. Among other things, this
enhancement may allow one to study in more generality to what extent
other theories of gravity are mapped onto the post-TOV
formalism. Another limitation of the formalism is that it is
intrinsically perturbative with respect to GR solutions. It is
important to generalize to theories of gravity that present screening
mechanisms; the viability of perturbative expansions in these theories
is a topic of active research (see
e.g.~\cite{Chagoya:2014fza,Avilez-Lopez:2015dja,Zhang:2016njn,McManus:2016kxu}).

We hope that the astrophysical applications outlined in this work will
stimulate more research to address these issues.

\begin{acknowledgments}
  K.G. is supported by the Ram\'{o}n y Cajal Programme of the Spanish
  Ministerio de Ciencia e Innovaci\'{o}n and by NewCompStar (a
  COST-funded Research Networking Programme). H.O.S., G.P. and
  E.B. are supported by NSF CAREER Grant No.  PHY-1055103. E.B. is
  supported by FCT contract IF/00797/2014/CP1214/CT0012 under the
  IF2014 Programme.  This work was supported by the
  H2020-MSCA-RISE-2015 Grant No. StronGrHEP-690904. H.O.S. thanks the
  Instituto Superior T\'ecnico for hospitality, and the Department of
  Physics and Astronomy of the University of Mississippi for financial
  support.
\end{acknowledgments}

\appendix


%

\end{document}